\documentstyle›12pt!{article}
\title{REGULAR UNIMODAL SYSTEMS AND FACTORS OF FINITE AUTOMATA}
   \author{Petr K\accent23urka \vspace{8pt}\\
   Faculty of Mathematics and Physics, Charles University\\
   Malostransk\'{e} n\'{a}m\v{e}st\'{\i} 25, 118 00 Praha 1,
 Czechia\\
    {\small fax: (422) 532 742, email: kurka@cspguk11.bitnet} }
\date{}  \ehyph

\newcommand{\mez}{\vspace{10pt}}
\newtheorem{dfn}{Definition}
\newtheorem{lem}{Lemma}
\newtheorem{pro}{Proposition}
\newtheorem{thm}{Theorem}
\newtheorem{cor}{Corollary}
\newtheorem{exm}{Example}

\begin{document}
\maketitle

\begin{abstract}
Dynamical systems at the edge of chaos, which have been considered
as models of self-organization phenomena, are marked by their
ability to perform nontrivial computations. To distinguish them
from systems with limited computing power, we formulate two
simplicity criteria for general dynamical systems, and apply them
to unimodal systems on real interval. We say that a dynamical
system is  regular, if it yields a regular language when observed
through arbitrary  almost disjoint cover. Finite automata are
regarded as dynamical systems  on zero-dimensional spaces and their
factors yield another class of  simple dynamical systems. These two
criteria coincide on subshifts,  since a subshift is regular iff
it is a factor of a finite automaton  (sofic systems).  A unimodal
system on real interval is regular if it has only a finite number
of periodic points, and nonrecursive otherwise. On the other hand
each  $S$-unimodal system with finite, periodic or preperiodic
kneading  sequence is a factor of a finite automaton. Thus
preperiodic  $S$-unimodal systems are factors of finite automata,
which are not regular. \mez \\
key words: edge of chaos, computing power, complex dynamics.

\end{abstract}
\pagebreak

\section{Introduction}

The complexity of a dynamical system might be characterized by the
regularity of its visits in different regions of the state space.
While the randomness of these visits is measured by statistical
properties like topological entropy, the study of their structural
properties requires the methods of computer science and formal
language theory. This approach is well established in the study of
$Z$-subshifts, which are in one-to-one correspondence with central
languages (see \cite{kn:Culik}). Regular $Z$-subshifts are called
sofic systems and are factors of subshifts of finite type.
In one-dimensional dynamics formal language methods have been
pioneered by Crutchfield and Young in \cite{kn:Crutchfield}, who
investigate the regularity of languages generated by logistic
system on standard partition of the state space.

In the present paper we generalize these results. Any cover of the
state space yields a language, and we are interested in its
regularity  and recursivity. To get a satisfactory theory we
restrict ourselves to covers, which are sufficiently simple.
Otherwise the complexity of the obtained language would reflect the
complexity of the cover, rather than the complexity of the
dynamics. In  zero-dimensional spaces the appropriate covers are
finite clopen partitions.   In the case of a compact real interval
we use interval covers consisting of closed intervals, which meet
at their endpoints. A general concept  is almost disjoint finite
cover. It consists of closed sets,  which have everywhere the  same
dimension equal to the dimension of the space, and meet  in sets
whose dimension is smaller. We say that a dynamical system is
regular (recursive), if it generates a regular (recursive) language
on each almost disjoint cover.

In formal language theory there is a duality between languages and
automata. In particular regular languages are recognized by finite
automata. Finite automata can be regarded as dynamical systems on
discontinuum, which is a compact, perfect, zero-dimensional space.
Dynamical systems on discontinuum are of particular interest, since
every dynamical system on a compact metric space is a factor of
some dynamical system on discontinuum  (generalized Alexandrov
Theorem, see \cite{kn:Balcar}). Every type of automata studied in
computer science corresponds in a natural way to a class of
dynamical systems on discontinuum. Since  factorization might only
lead to simplification, the hierarchy of these automata induces a
hierarchy of dynamical systems. Factors  of finite automata are at
the bottom of this hierarchy.

We show that the simplest dynamical systems, which have only finite
attractors, are both regular and factors of finite automata.
Equivalence between these two classes occurs also in subshifts.
Since we work with noninvertible dynamical systems, the languages
we get are in general not central, but only right central, i.e.
extendable to the right. Right central languages correspond to
$N$-subshifts in a similar manner as central languages correspond
to $Z$-subshifts (see Culik II et al. \cite{kn:Culik}).
An $N$-subshift is regular iff it is a factor of a finite
automaton. More generally, any zero-dimensional factor of a finite
automaton is regular, but not vice versa.

For unimodal systems on real interval we get a reverse situation:
If we restrict ourselves to the standard cover consisting of the
two intervals meeting at the turning point, we get a regular
language iff the kneading sequence is either finite, or periodic
or preperiodic (this is the generalization of results of
\cite{kn:Crutchfield}). However, the interval covers with cut
points whose itineraries are nonrecursive, yield nonrecursive
languages. Such covers occur whenever there exists an infinite
number of periodic points. Thus a unimodal system is regular if it
has a finite number of periodic points, and nonrecursive otherwise.

These results have been obtained without any assumptions on
smoothness.  To investigate  unimodal factors of finite automata,
we need a stronger  theory of $S$-unimodal systems (negative
Schwarzian derivative). We have proved in \cite{kn:Kurka} that any
$S$-unimodal system whose kneading sequence is either preperiodic
or periodic odd is a factor of a finite automaton. On the other
hand the $S$-unimodal systems  with aperiodic kneading sequences,
which occur at the common limit of period doubling and band merging
bifurcations, are not factors of finite automata. In the present
paper we strengthen the first result and show that any $S$-unimodal
system with finite, periodic or preperiodic kneading sequence is
a factor of a finite automaton. Thus preperiodic $S$-unimodal
systems are factors of finite automata, which are not regular.

\section{Dynamical systems and languages}

We conceive dynamical system as a self map of a topological space,
which is compact, metrizable and homogenous, i.e. has at each point
the same dimension. Recall $dim(X) = -1$ if $X = 0$,
$dim(X)= \sup \{dim_{x}(X) º x \in X \}$ if $X \neq 0$,
$dim_{x}(X) \leq n+1$ if there exist arbitrary small open
neighbourhoods of $x$ whose boundaries have dimension at most $n$,
$dim_{x}(X) = n+1$ if $dim_{x}(X) \leq n+1$ and $dim_{x}(X)
\not\leq n$, $dim_{x}(X)=\infty$, if $(\forall n)(dim_{x}(X)
\not\leq n)$.  A cover of space $X$ is a system $\alpha = \{V_{a}º
a \in A\}$ of its subsets whose union is $X$. A cover $\alpha$ is
finer than $\beta = \{V_{b}º b \in B \}$ ($\alpha \geq \beta$),
if $(\forall a \in A)(\exists b \in B)(V_{a} \subseteq V_{b})$.

\begin{dfn}
A space $X$ is homogenous, if $(\forall x \in X)(dim_{x}(X) =
dim(X))$.  A cover $\alpha$ of $X$ is almost disjoint if it is
finite,  each $V_{a}$ is a closed homogenous set with $dim(V_{a})
= dim(X)$,  and  $dim(V_{a} \cap V_{a'}) < dim(X)$ for $a \neq a'$.
\end{dfn}

Every zero-dimensional space is homogenous, and a real interval is a
homogenous one-dimensional space.
A cover of a zero-dimensional compact space is almost disjoint iff it
is a finite clopen partition, i.e. if it consists of disjoint
clopen (closed and open) sets. A cover $\alpha = \{V_{a} º a \in
A\}$ of a compact real interval is almost disjoint, iff each
$V_{a}$ is  a finite union of closed intervals, and $V_{a} \cap
V_{b}$ consists  of endpoints of these intervals. We say that
$\alpha$ is an interval cover, if each $V_{a}$ is a closed
interval. The endpoints of these intervals are referred to as
cutpoints of $\alpha$. A space is perfect, if each its point is a
limit point. A discontinuum is a compact metrizable space, which
is 0-dimensional and perfect. Any two discontinua are homeomorphic,
and any countable product of  discrete finite spaces is a
discontinuum  (see e.g. \cite{kn:Hocking}).

\begin{dfn}
A dynamical system $(X,f)$ is a continuous map $f: X \rightarrow
X$ of a compact metrizable and homogenous space $X$ to itself.
A homomorphism $\varphi: (X,f) \rightarrow (Y,g)$ of dynamical
systems is a continuous mapping $\varphi:X \rightarrow Y$ such that
$g \varphi = \varphi f$. We say that $(Y,g)$ is a factor of
$(X,f)$, if $\varphi$ is a factorization (i.e. a surjective map).
\end{dfn}

The $n$-th iteration of a dynamical system $(X,f)$ is defined by
$f^{0}(x)=x$, $f^{n+1}(x)=f(f^{n}(x))$. A point $x \in X$ is
periodic  with period $n>0$, if $f^{n}(x)=x$ and $f^{i}(x) \neq x$
for $0<i<n$. A point $x$ is eventually periodic, if there exists
$i \geq 0$, such that $f^{i}(x)$ is periodic. It is preperiodic,
if it is eventually periodic but not periodic. It is aperiodic, if
it is not eventually periodic. A subset $A \subseteq X$ is
invariant, if $f(A) \subseteq  A$, and strongly invariant if
$f(A)=A$.  The $\omega$-limit set of a set $A \subseteq  X$ is
$\omega(A)= \bigcap_{n} \overline{ \bigcup_{m>n} f^{m}(A)}$. If $A$
is nonempty, then $\omega(A)$ is a nonempty, closed, strongly
invariant set. If $\varphi: (X,f) \rightarrow (Y,g)$ is a
factorization, then $\varphi(\omega_{f}(A)) =
\omega_{g}(\varphi(A))$.

Let $A$ be a finite alphabet, and $N = \{0,1,2,...\}$  the set of
non-negative integers. We frequently use alphabet $2=\{0,1\}$.
Denote $A^{N}= \{u = u_{0}u_{1}... º u_{i} \in A \}$ the space of
infinite sequences of letters of $A$, with product topology. For
$n \in N$ denote $A^{n}$ the set of sequences of length $n$, $A^{*}
= \cup_{n \in N} A^{n}$ the set of finite sequences, and
$\overline{A^{*}} = A^{*} \cup A^{N}$. Denote $ºuº$ the length of
a sequence $u \in \overline{A^{*}}$, ($0 \leq ºuº \leq \infty)$,
and $\lambda$ the word of zero length so, $u_{i}=\lambda$ for
$i>ºuº$.  Write $u \unlhd v$ if $u$ is a  substring of $v$, and $u
\sqsubseteq v$,  if $u$ is an initial  substring of $v$. Denote
$u_{ºi} = u_{0}...u_{i-1}$the initial substring of $u$ of length
$i$. If $u \in A^{*}$, denote $\overline{u} \in A^{N}$ the infinite
repetition of $u$, defined by $\overline{u}_{kºuº+i} = u_{i}$. For
$u \in \overline{A^{*}}$ define $\sigma(u) \in \overline{A^{*}}$
by $\sigma(u)_{i} = u_{i+1}$. Thus $\sigma(\lambda) = \lambda$. For
$u \in A^{n}$ denote $›u! = \{v \in A^{N}º v_{ºn} = u \} \subseteq
A^{N}$ the cylinder determined by $u$. It is a clopen  set, and
$\alpha_{n} = \{›u! º u \in A^{n}\}$ is a clopen partition of
$A^{N}$. A subshift is any nonempty subset $\Sigma \subseteq A^{N}$,
which is closed and $\sigma$-invariant. If $\Sigma$ is a subshift,
then $(\Sigma,\sigma)$ is a dynamical system.

\begin{dfn}
Let $(X,f)$ be a dynamical system, and $\alpha= \{V_{a}º a \in A\}$
a finite cover of $X$. The language and subshift generated by
$(X,f)$ on $\alpha$ are
\› {\cal L}(X,f,\alpha) = \{ u \in A^{*}º V_{u} \neq 0 \}, \;\;\;
 {\cal S}(X,f,\alpha) = \{ u \in A^{N}º V_{u} \neq 0 \}, \!
where $V_{u} = \{x \in X º (\forall i<ºuº)(f^{i}(x) \in V_{u_{i}})
\}$ is the set of points, which visit sets of $\alpha$ according
to  the sequence $u$. We say that $(X,f)$ is regular (recursive),
if ${\cal L}(X,f,\alpha)$  is regular (recursive) for every almost
disjoint cover $\alpha$ of $X$.
\end{dfn}

\begin{dfn}
A nonempty language $L \subseteq  A^{*}$ is right central, if \\
$(\forall u \in L)(\forall v \unlhd u)(v \in L)$, and \\
$(\forall u \in L)(\exists a \in A)(ua \in L)$.\\
The adherence of $L \subseteq A^{*}$ and the language of
$\Sigma \subseteq A^{N}$ are defined by\\
${\cal A}(L) = \{u \in A^{N}º (\forall v \in A^{*})
      (v \unlhd u \Rightarrow v \in L ) \}$.\\
${\cal L}(\Sigma)$ = $\{u \in A^{*}º(\exists v \in \Sigma)(u \unlhd
v)\}$ = ${\cal L}(\Sigma, \sigma, \{ ›a! º a \in A  \})$.
\end{dfn}

If $(X,f)$ is a dynamical system and $\alpha$ a closed cover of
$X$, then ${\cal L}(X,f,\alpha)$ is a right central language,
${\cal S}(X,f,\alpha)$ is a subshift, ${\cal L}({\cal S}
(X,f,\alpha))$ = ${\cal L}(X,f,\alpha)$,  and ${\cal A}
({\cal L}(X,f,\alpha)) = {\cal S}(X,f,\alpha)$.  If $L \subseteq
A^{*}$ is a right central language and $\Sigma \subseteq A^{N}$ a
subshift, then ${\cal A}(L)$ is a subshift, ${\cal L}(\Sigma)$ is
a right central language, ${\cal A}{\cal L}(\Sigma) = \Sigma$, and
${\cal L}{\cal A}(L) = L$. This can be proved by a simple
compactness argument (cf. \cite{kn:Culik}).

\begin{pro} \label{refinement}
Let $(X,f)$ be a dynamical system and $\alpha \geq \beta$ almost
disjoint covers of $X$. If ${\cal L}(X,f,\alpha)$ is regular
(recursive), then so is ${\cal L}(X,f,\beta)$.
\end{pro}
Proof: Suppose that there exist $a \in A$, and $b \neq b' \in B$
with $V_{a} \subseteq V_{b} \cap V_{b'}$. Then $dim(V_{a}) \leq
dim(V_{b} \cap V_{b'}) < dim(X)$, which is a contradiction. Thus
for each $a \in A$ there exists a unique $h(a) \in B$ with $V_{a}
\subseteq  V_{h(a)}$. We prove $V_{b} = \cup \{V_{a}º h(a) = b \}$.
If not, denote $Y=V_{b} - \cup \{V_{a}º h(a)=b\}$. For each $y \in
Y$ there exist $a'$, $b'$, with $y \in V_{a'} \subseteq V_{b'}$,
thus $Y \subseteq \cup\{V_{b} \cap V_{b'}º b' \neq b\}$, and
$dim(Y) < dim(V_{b})$.  Since $Y$ is open in $V_{b}$, $V_{b}$ is
not homogenous, which is a contradiction. The map $h$ can be
extended to a monoid homomorphism $h: A^{*} \rightarrow B^{*}$,
which is $\lambda$-free, i.e. $h^{-1}(\lambda) = \lambda$. We get
${\cal L}(X,f,\beta) = h({\cal L}(X,f,\alpha))$, and since both
regular and recursive languages are closed to $\lambda$-free
homomorphisms (see \cite{kn:Hopcroft}),  ${\cal L}(X,f,\beta)$  is
regular (recursive). $\Box$

\begin{pro} \label{regular0}
A dynamical system $(X,f)$ on a zero-dimensional space $X$ is regular
(recursive) iff there exists a sequence $\alpha_{n}$ of clopen
partitions of $X$  such that $\alpha_{n+1} \geq \alpha_{n}$,
$diam(\alpha_{n})  \rightarrow 0$, and  ${\cal L}(X,f,\alpha_{n})$
is regular (recursive).
\end{pro}
Proof: Let $\beta$ be a finite clopen partition. For each $x \in
X$ there exists $n_{x}$ and $A_{x} \in \alpha_{n_{x}}$ with $x \in
A_{x}$, such that $\{A_{x}º x \in X\} \geq \beta$. Let
$\{A_{x}º x \in K \}$ be a finite subcover of $\{A_{x}º x \in X\}$,
and $n = \max \{n_{x} º x \in K \}$. Then $\alpha_{n} \geq \beta$,
and we get the result by Proposition \ref{refinement}. $\Box$ \mez

\begin{pro} \label{regular1}
A dynamical system $(I,g)$ on a compact real interval is regular
(recursive) iff ${\cal L}(I,g,\alpha)$ is regular (recursive)  for
each interval cover $\alpha$.
\end{pro}
Proof: If $\alpha$ is an almost disjoint cover of $I$, then each
$V_{a}$ is a finite union of closed nontrivial intervals, so there
exists an interval cover, which refines it, and we apply
Proposition  \ref{refinement}. $\Box$

\begin{dfn}
A set $A \subseteq  X$ is an attractor, of a dynamical system
$(X,f)$, if it has a closed neighbourhood $U$ such that $f(U)
\subseteq int(U)$,   and $A=\omega(U)$ (see \cite{kn:Devaney}). We
say that $(X,f)$  is a system with finite attractors, if all its
attractors are finite,  i.e. if $\omega(X)$ is finite .
\end{dfn}

\begin{lem} \label{attractor}
Any attractor is stable, i.e. any its neighbourhood contains an
invariant neighbourhood.
\end{lem}
Proof: Let $U \subseteq  X$ be closed neighbourhood of $A$,  $f(U)
\subseteq  int(U)$, and $A= \omega(U)$. Let $V \subseteq  U$  be
an open neighbourhood of $A$. Suppose that $V$ does not contain any
invariant neighbourhood of $A$. For $k>0$ denote $W_{k} = \cup_{n
\in N} f^{n}(B_{1/k}(A))$,  where $B_{\varepsilon}(A) = \{x \in Xº
\varrho(y,A) < \varepsilon\}$. $W_{k}$ is an invariant
neighbourhood of $A$, so $W_{k} \not \subseteq V$. Thus there
exists some $x_{k} \in B_{1/k}(A)$, and an integer $n_{k}$ with
$f^{n_{k}}(x_{k}) \in X-V$. Since $X-V$ is compact,  there exists
a subsequence $k_{i}$, with $f^{n_{k_{i}}}(x_{k_{i}}) \rightarrow
x \in X-V$. For all sufficiently large $k$, $x_{k} \in U$, so $x
\in \omega(U)=A$, which is a contradiction. $\Box$

\begin{thm} \label{regular attractor}
Any dynamical system with finite attractors is regular.
\end{thm}
Proof: Let $(X,f)$ be a dynamical system with  finite $\omega(X)$.
Since $\omega(X)$ is strongly invariant, it consists of the
periodic points of $X$. Let $\alpha=\{V_{a} º a \in A\}$ be  an
almost disjoint  cover of $X$. By Lemma
\ref{attractor} there exists a system $\{U_{p} º p \in \omega(X)
\}$, such that  $U_{p}$ is a neighbourhood of $p$, if $p \not\in
V_{a}$, then  $U_{p} \cap V_{a} = 0$, and $U = \cup_{p \in
\omega(X)} U_{p}$ is invariant. Denote
$L = \{u \in A^{*} º (\exists p \in \omega(X))(\forall i < ºuº)
(f^{i}(p) \in V_{u_{i}}) \}$. Then $L$ is a regular language. For
each $x \in X$ there exists  $n_{x}$ with $f^{n_{x}}(x) \in U$, and
open neighbourhood $W_{x}$  of $x$ with $f^{n_{x}}(W_{x}) \subseteq
U$. Since $X$ is compact,  there exists a finite subcover
$\{W_{x}ºx \in K\}$. Let $n = \max\{n_{x}ºx \in K\}$. Then for any
$u \in {\cal L}(X,f,\alpha)$, $\sigma^{n}(u) \in L$. It follows
that ${\cal L}(X,f,\alpha)$ is a regular language. $\Box$

\section{Finite automata}

A multitape Turing machine is a finite control device connected to
several potentially infinite tapes. We can assume that the control
device itself occupies a field on a tape, which does not move. The
resulting machine is conceptually simpler. It consists of a finite
number of doubly infinite tapes, which interact at the scanned
positions. The contents of the scanned positions of all tapes
determine the move of the machine: the letters which it writes on
the tapes and the directions of their shifts. In contrast to
computer science we assume that the tapes are actually infinite
and have also infinite content. Thus we obtain a dynamical  system
on discontinuum (cf. \cite{kn:Moore}). In a similar manner we
conceive a multitape finite automaton. It is a finite control
attached to input tapes, on which all future inputs are written.
Moreover there are output tapes, to which the results are written.
Thus the tapes move in one direction only.

Let $A$ be a finite alphabet and denote $Z = \{...-1,0,1,...\}$
the set of integers. A tape is a data structure containing the
letters of $A$ indexed by $Z$. The state of a tape is a map $u: Z
\rightarrow A$, thus its state space is $A^{Z}$, which is a
discontinuum. The tape state may be updated by rewriting the zeroth
position and shifting the tape left or right. This is accomplished
by (continuous) updating functions $\sigma_{a}^{i}: A^{Z}
\rightarrow A^{Z}$ defined by $\sigma_{a}^{i}(u)_{j} = a$
if $i+j = 0$, and $\sigma_{a}^{i}(u)_{j} = u_{i+j}$ otherwise.
Note that $\sigma^{i}(u) = \sigma^{i}_{u_{0}}(u)$ coincides with
the previous use.
\› \begin{array}{rcl}
  \sigma_{a}^{-1}(...u_{-2}u_{-1}u_{0}u_{1}u_{2}...) & =
             & ...u_{-3}u_{-2}u_{-1}au_{1}...\\
  \sigma_{a}^{0}(...u_{-2}u_{-1}u_{0}u_{1}u_{2}...) & =
             & ...u_{-2}u_{-1}au_{1}u_{2}...\\
  \sigma_{a}^{1}(...u_{-2}u_{-1}u_{0}u_{1}u_{2}...) & =
             & ...u_{-1}au_{1}u_{2}u_{3}... \end{array} \!

\begin{dfn}
A Turing automaton is a finite system of finite alphabets
$(A_{t})_{t \in T}$ together with transition functions
$\delta_{t}: Q \rightarrow A_{t}$, $\eta_{t}: Q \rightarrow
\{-1,0,1\}$, where $Q = \prod_{t \in T} A_{t}$ is the set of
inner states. A Turing automaton determines a dynamical system
$(X,f)$, where $X=\prod_{t \in T} A_{t}^{Z}$, $\pi: X \rightarrow
Q$ is the projection $\pi(u)_{t} = u_{t0}$, and
\› f(u)_{t} = \sigma_{a}^{i}(u_{t})\; \mbox{ where } \;
   a=\delta_{t}\pi(u),\; i=\eta_{t}\pi(u),\;\;\; u \in X, t \in
T.\!
\end{dfn}
The advance of a tape $t \in T$ on $u \in X$ in $n$ steps is
$d_{t}(u,n) = \sum_{i=0}^{n-1} \eta_{t}\pi f^{i}(x)$. A finite
automaton is a Turing automaton, whose all tapes move to the left,
i.e. $(\forall t \in T)(\forall q \in Q)(\eta_{t}(q) \geq 0)$.

\begin{thm} \label{finiteattractor}
A dynamical system with finite attractors is a factor of a finite
automaton.
\end{thm}
Proof: Let $(Y,g)$ be a system with finite attractors. Then
$\omega(Y)$ consists of periodic points. Denote $t_{p}$ the period
of $p \in \omega(Y)$. By Lemma \ref{attractor} there exists a
$g^{t_{p}}$-invariant neighbourhood $U$ of $p \in \omega(Y)$,
which does not meet $\omega(Y)-\{p\}$. It follows that
$Y_{p} = \{y \in Yº g^{kt_{p}}(y) \rightarrow p \}$ is an open set,
and  $\{Y_{p} º p \in \omega(Y) \}$ is a clopen partition of $Y$.
Construct a two-tape finite automaton $(X,f)$ with alphabets $A_{1}
=  A_{2} = \omega(Y)$. Define transition functions by
$\eta_{1}(p,q)  = 0$, $\eta_{2}(p,q) = 1$, $\delta_{1}(p,q) =
\delta_{2}(p,q) = g(p)$.  Thus the first tape is stationary and
only one of its fields  is used. Its content changes according to
$g$, and the results are written to the second (output) tape. As
the time proceeds, the  second tape contains ever longer periodic
segments. For $p \in \omega(Y)$, $k \geq 0$ define $X_{pk}
= \{(u,v) \in X º u_{0}=p \; \& \; \min \{j > 0 º g^{j}(v_{-j})
\neq u_{0}\} = k+1 \}$, $X_{p\infty} = \{u \in X º (\forall j > 0)
(g(v_{-j})  =  u_{0} = p) \}$. Then $f(X_{pk}) = X_{g(p),k+1}$. For
$k \geq 0$ define $\pi_{k}: X \rightarrow 2^{N}$ by $\pi_{k}(u,v)
= v_{-k-1}v_{-k-2}...$ . Then $\pi_{k+1}f (u,v) = \pi_{k}(u,v)$ for
$(u,v) \in X_{pk}$. Let $Y_{p0} = Y_{p} \cap   \overline{Y-g(Y)}$.
For $p \in \omega(Y)$ there exists a mapping $\psi_{p}: 2^{N}
\rightarrow Y$ such that $\psi_{p}(2^{N}) = Y_{p0}$ if $Y_{p0} \neq
0$, and $\psi_{p}(2^{N}) = \{p\}$ if $Y_{p0} = 0$. For $(u,v) \in
X_{pk}$ define $\varphi(u,v) = g^{k}\psi_{g^{-k}(p)}\pi_{k}(u,v)$.
For $(u,v) \in X_{p\infty}$ define $\varphi(u,v)=p$. For any  $y
\in Y -  \omega(Y)$ there exists only a finite chain of preimages,
so $\varphi$ is surjective and it is clearly continuous. If $(u,v)
\in X_{pk}$, then $\varphi f(u,v) = g^{k+1}\psi_{g^{-k-1}(g(p))}
\pi_{k+1} f(u,v)$  = $g^{k+1}\psi_{g^{-k}(p)} \pi_{k}(u,v)$ = $g
\varphi (u,v)$.  If $(u,v) \in X_{p\infty}$, then $\varphi f(u,v)
= g(p) = g \varphi (u,v)$.  Thus $\varphi: (X,f) \rightarrow (Y,g)$
is a factorization. $\Box$

\begin{thm} \label{sizeper}
Let $(X,f)$ be a factor of a finite automaton. Then there is an
integer $q>0$ such that each nonempty open set $U \subseteq X$
contains a point $x \in U$ such that $\omega(x)$ has at most $q$
elements.
\end{thm}
Proof: The property is preserved by factorization, so it suffices
to prove it for finite automata. Let $(X,f)$ be a finite automaton
and denote $q$ the size of $Q$. Let $U \subseteq  X$ be an open
set and $x \in U$. There exists $n$ with $›x_{ºn}! \subseteq U$,
where $›x_{ºn}! = \{u \in Xº (\forall t \in T)(\forall ºiº \leq
n)(u_{ti} =  x_{ti}) \}$. Denote $T_{x} = \{ t \in T º \lim_{k
\rightarrow \infty} d_{t}(x,k) = \infty \}$ ($T_{x}$ might be
empty). There exist $m$, and $k \leq q$, such that $(\forall  t \in
T-T_{x})(\forall i)(d_{t}(f^{m}(x),i)=0)$, $(\forall t \in
T_{x})(d_{t}(x,m) > n)$, $\pi f^{m}(x) = \pi f^{m+k}(x)$. Denote
$d_{t0}=d_{t}(x,m)$, $d_{t1}=d_{t}(x,m+k)$ and define a point $y
\in X$ by\\
$y_{ti} = x_{ti}$ if $t \in T-T_{x}$ \\
$y_{ti} = x_{ti}$ if $t \in T_{x}$, $i \leq d_{t1}$,\\
$y_{ti} = x_{tj}$ if $t \in T_{x}$, $i> d_{t1}$, $d_{t0} < j \leq
d_{t1}$, and $d_{1}-d_{0}$ divides $i-j$.\\
Then $y \in ›x_{ºn}!$, and $g^{i}(y)$ converges to a periodic point
with period $k \leq q$. $\Box$

\begin{pro} \label{regularfactor}
Any zero-dimensional factor of a finite automaton is regular.
\end{pro}
Proof: If $\varphi: (X,f) \rightarrow (Y,g)$ is a factorization,
and if $\alpha$ a clopen partition of $Y$, then
$\varphi^{-1}(\alpha) =  \{\varphi^{-1}(A)º A \in \alpha \}$ is a
clopen partition of $X$  and ${\cal L}(Y,g,\alpha)$ =
${\cal  L}(X,f,\varphi^{-1}(\alpha))$. $\Box$

\begin{thm} \label{regularshift}
A subshift is regular iff it is a factor of a finite automaton.
\end{thm}
Proof: Let $(X,f)$ be a finite automaton and let $\alpha_{n} =
\{V_{v} º v \in A \}$ be the clopen partition of $n$-cylinders
$V_{v} = \{u \in Xº (\forall t \in T)(\forall ºiº \leq n)
(u_{ti} = v_{ti})\}$ (here $A= \prod_{t \in T} A_{t}^{2n+1}$ ).
Let us prove that if
$V_{u_{0}...u_{m-1}} \neq 0$, and $V_{u_{m-1}u_{m}} \neq 0$, then
$V_{u_{0}...u_{m}} \neq 0$. Choose $x \in V_{u_{0}...u_{m-1}}$, $y
\in V_{u_{m-1}u_{m}}$, and define\\
$z_{ti} = x_{ti}$ if $i \leq d_{t}(x,m)+n$,\\
$z_{ti} = y_{tj}$ if $i = d_{t}(x,m)+n+1$, $j=d_{t}(y,1)+n$.\\ Then
$z \in V_{u_{0}...u_{n}}$. Thus ${\cal L}(X,f,\alpha_{n})$ is
regular (it is a finite complement language) and $(X,f)$ is regular
by Proposition \ref{regular0}. By Proposition \ref{regularfactor}
any subshift which is a factor of a finite automaton is regular.
On the other hand if $\Sigma \subseteq A^{N}$ is a regular
subshift, then its language is recognized by a finite  accepting
automaton, so there exists a set of states $Q$,  initial set $q_{0}
\in Q$, accepting set $Q_{1} \subseteq  Q$,  and a transition
function $\delta: Q \times A \rightarrow Q$ such that $u \in {\cal
L}(\Sigma)$ iff $\delta^{*}(q_{0},u) \in Q_{1}$. Since ${\cal
L}(\Sigma)$ is right central, $q_{0} \in Q_{1}$ and for any $q \in
Q_{1}$ there exists some $h(q) \in A$ with $\delta(q,h(q)) \in
Q_{1}$. Define a two-tape automaton with alphabets $A_{1}=Q_{1}$,
$A_{2}=A$, $\eta_{1}(q)=0$, $\eta_{2}(q)=1$ for any $q \in Q_{1}$.
The transitions are defined by
\› \begin{array}{lcll}
   \delta(q,a) \in Q_{1} & \Rightarrow & \delta_{1}(q,a) =
\delta(q,a), & \delta_{2}(q,a)=a,\\
   \delta(q,a) \not\in Q_{1} & \Rightarrow &
\delta_{1}(q,a)=\delta(q,h(q)), & \delta_{2}(q,a)=h(q)
\end{array} \!
Define $\varphi: Q_{1}^{Z} \times  A^{Z} \rightarrow A$ by
$\varphi(u,v) = \delta_{2}(u_{0},v_{0})$. then $\varphi:(X,f)
\rightarrow (\Sigma,\sigma)$ is a factorization. $\Box$ \mez

\noindent
Since the set of the inverse words of a regular language is
regular, any $Z$-subshift of a finite type is a factor of
a finite automaton.
Since a closed hyperbolic set with a local product structure (see
Shub \cite{kn:Shub}) is a factor of a $Z$-subshift of finite type,
it is also a factor of a finite automaton.

\begin{exm}
There exists a regular system on $0$-dimensional space, which is
not a factor of a finite automaton.
\end{exm}
Proof: Take $X = 2^{N}$ and for $n>0$ denote $U_{n} = ›0^{n-1}1!$.
For each $n>0$ let $\{U_{nm}º 0\leq m<n\}$ be a clopen partition
of $U_{n}$. Define $g$ so that if $x \in U_{n,m}$, then $g(x) \in
U_{n,m+1 \mbox{ mod } n}$, and $g^{n}(x)=x$. Thus each $x \in
U_{n}$ is a periodic point with period $n$. Then $(X,g)$ is a
regular system, which is by Theorem \ref{sizeper} not a factor of
a finite
automaton. $\Box$

\section{Unimodal systems}

\begin{dfn}
A dynamical system $(I,g)$ defined on a real interval $I=›a,b!$ is
unimodal, if $g(a)=g(b)=a$, and if there exists a turning point $c
\in (a,b)$, such that $g$ is increasing in $›a,c!$, and
decreasing in $›c,b!$.
\end{dfn}

A powerful tool for investigating unimodal systems is the kneading
theory, which is based on the concept of itineraries generated by
the partition $\{›a,c),\{c\},(c,b!\}$. We modify this theory using
the cover $\{›a,c!,›c,b!\}$, or more general interval covers
instead. We lose the uniqueness of the itinerary, but the resulting
theory is simpler. In particular a simple compactness argument
yields in  Theorem \ref{regularstandard} a necessary and sufficient
condition characterizing the language generated by the standard
cover, instead of only sufficient
condition in Theorem 18.12 of \cite{kn:Devaney}. Moreover, we do
not need any differentiability assumptions.

Throughout this section we assume that $\alpha$ is an interval
cover, which contains among its cut points the turning point $c$.
Thus let $a_{0}<...<a_{n}$ be an increasing sequence such that
$a_{0}$, $a_{n}$ are the endpoints of $I$, and $a_{m}=c$ is the
turning point, where $0<m<n$. Denote
$V_{j}=›a_{j},a_{j+1}!$, and $\alpha = \{V_{j}º j \in A \}$, where
$A=\{0,...,n-1\}$. For $j \in A$ denote $\tau(j)=0$ if $j<m$,
$\tau(j)=1$ if $j \geq m$. For $u \in A^{*}$, $k \leq ºuº$, denote
$\tau_{k}(u) = \sum_{i=0}^{k-1} \tau(u_{i}) \mbox{ mod } 2$. We say
that $u \in A^{*}$ is even (odd), if $\tau_{ºuº}(u)$ is zero (one).

\begin{dfn}
For $u,v \in \overline{A^{*}}$, define $u \prec v$ iff there exists
$k$ such that $u_{ºk}=v_{ºk}$, and either $\tau(u_{ºk})=0$ and
$u_{k}<v_{k}$, or $\tau(u_{ºk})=1$ and $u_{k} > v_{k}$. (This
includes the case $ºuº=k$ or $ºvº=k$, when $u_{k} = \lambda$ or
$v_{k} = \lambda$. We put in this case $0<...<m-1 < \lambda < m <
... < n-1$.)  Define $u \preceq v$ iff either $u \prec v$, or $u
\sqsubseteq v$  or $v \sqsubseteq u$.
\end{dfn}

For example if $n=2$, then we get $00 \prec 0 \prec 01 \prec
\lambda \prec 11 \prec 1 \prec 10$. If $v \not\preceq u$ then $u
\prec v$, and $u \preceq v$. If $v \not\prec u$, then $v \preceq
u$. Thus $\prec$ is a strict ordering on $\overline{A^{*}}$, and
$\preceq$ is an ordering on $A^{N}$.

\begin{pro} \label{prec}
If $u,v \in \overline{A^{*}}$, and $u \prec v$, then
$(\forall x \in V_{u}) (\forall y \in V_{v})(x \leq y)$.
\end{pro}
Proof: Let $u \prec v$, $k<\min(ºuº,ºvº)$, $u_{ºk} = v_{ºk}$,
$u_{k} \neq v_{k}$, $x \in V_{u}$, $y \in V_{v}$. We prove $x \leq
y$ by induction on $k$. If $k=0$, then $u_{0} < v_{0}$, and $x \leq
y$. If $k>0$, we apply the Proposition to $\sigma(u)$ and
$\sigma(v)$. If
$\tau(u_{0})=\tau(v_{0})=0$, then $\sigma(u) \prec \sigma(v)$, so
$g(x) \leq g(y)$, and $x \leq y$ since $x,y \leq c$. If
$\tau(u_{0})=\tau(v_{0})=1$, then $\sigma(u) \succ \sigma(v)$, so
$g(x) \geq g(y)$, and $x \leq y$ since $x,y \geq c$.  $\Box$

\begin{dfn}
The upper and lower itineraries of $x \in I$ are defined by
\› {\cal I}^{\alpha}(x)  =  \max \{u \in A^{N}º x \in V_{u}\},\;\;\;
{\cal I}_{\alpha}(x)  =  \min \{u \in A^{N}º x \in V_{u}\} \!
(max and min are meant with respect to $\preceq$).
For $w \in A^{N}$, $j \in A$ denote
\› \begin{array}{l}
   L^{j}(w) = \{ u \in \overline{A^{*}}º (\forall i)(u_{i} = j
          \Rightarrow \sigma^{i+1}(u) \preceq w) \} \\
   L_{j}(w) = \{ u \in \overline{A^{*}}º (\forall i)(u_{i} = j
          \Rightarrow   w \preceq \sigma^{i+1}(u)) \}
\end{array} \!
\end{dfn}
The upper and lower itineraries might be obtained by inductive
definition:
\› \begin{array}{ccl}
a_{j} < g^{i}(x) < a_{j+1} & \Rightarrow & {\cal I}^{\alpha}(x)_{i}
= {\cal I}_{\alpha}(x)_{i} = j\\
g^{i}(x) = a_{j},\;\; \tau({\cal I}^{\alpha}(x)_{ºi}) = 0
   & \Rightarrow & {\cal I}^{\alpha}(x)_{i} = \min\{j,n-1\} \\
g^{i}(x) = a_{j},\;\; \tau({\cal I}^{\alpha}(x)_{ºi}) = 1
   & \Rightarrow & {\cal I}^{\alpha}(x)_{i} = \max\{j-1,0\} \\
g^{i}(x) = a_{j},\;\; \tau({\cal I}_{\alpha}(x)_{ºi}) = 0
   & \Rightarrow & {\cal I}_{\alpha}(x)_{i} = \max\{j-1,0\} \\
g^{i}(x) = a_{j},\;\; \tau({\cal I}_{\alpha}(x)_{ºi}) = 1
   & \Rightarrow & {\cal I}_{\alpha}(x)_{i} = \min\{j,n-1\}
\end{array} \!

\begin{pro}  \label{uplow}
$x \in V_{{\cal I}^{\alpha}(x)} \cap V_{{\cal
I}_{\alpha}(x)}$. If $x<y$, then ${\cal I}^{\alpha}(x) \preceq
{\cal I}_{\alpha}(y)$.
\end{pro}
Proof: If  ${\cal I}^{\alpha}(x) \succ {\cal I}_{\alpha}(y)$, then
$x \geq y$ by Proposition \ref{prec}.

\begin{pro} \label{unilanguage}
If $u \in \overline{A^{*}}$, then $u \in {\cal L}(I,g,\alpha) \cup
{\cal S}(I,g,\alpha)$ iff \\
$u \in L_{j}({\cal I}_{\alpha}(g(a_{j}))) \cap
       L^{j}({\cal I}^{\alpha}(g(a_{j+1})))$ for each $j<m$, and
\\ $u \in L^{j}({\cal I}^{\alpha}(g(a_{j}))) \cap
       L_{j}({\cal I}_{\alpha}(g(a_{j+1})))$ for each $j \geq m$.
\end{pro}
Proof: Let $u \in {\cal L}(I,g,\alpha)
\cup {\cal S}(I,g,\alpha)$, and $u_{i}=j$. Pick some  $x \in
V_{u}$. Then $g^{i}(x) \in V_{\sigma^{i}(u)} \subseteq V_{j}$. If
$j<m$, then $g(a_{j}) \leq g^{i+1}(x) \leq g(a_{j+1})$. Suppose
that ${\cal I}_{\alpha}(g(a_{j})) \succ \sigma^{i+1}(u)$. Then
$g(a_{j}) \geq g^{i+1}(x)$ by Proposition \ref{prec}, so $g(a_{j})
= g^{i+1}(x)  \in V_{\sigma^{i+1}(u)}$. This is however impossible,
as ${\cal I}_{\alpha}(g(a_{j}))$ is the least sequence $v$, for
which $g(a_{j}) \in V_{v}$. Thus ${\cal I}_{\alpha}(g(a_{j}))
\preceq \sigma^{i+1}(u)$, and $u \in L_{j}({\cal
I}_{\alpha}(g(a_{j})))$. Similarly we prove the other inequalities,
so the condition is necessary. To prove the sufficiency, we proceed
by induction on the length of $u$. If $ºuº \leq 1$, then $V_{u}
\neq 0$. If $ºuº$ is finite and the Proposition holds for $ºuº-1$,
we apply it to $\sigma(u)$. Thus there exists some $y \in
V_{\sigma(u)}$. If $u_{0}=j<m$, then ${\cal I}_{\alpha}(g(a_{j}))
\preceq \sigma(u) \preceq {\cal I}^{\alpha}(g(a_{j+1}))$. If ${\cal
I}_{\alpha}(g(a_{j})) \not\prec \sigma(u)$, then ${\cal
I}_{\alpha}(g(a_{j})) \sqsupseteq \sigma(u)$, and $a_{j} \in V_{u}
\neq 0$. Similarly if $\sigma(u) \not\prec {\cal
I}^{\alpha}(g(a_{j+1}))$, then $a_{j+1} \in V_{u}$. If ${\cal
I}_{\alpha}(g(a_{j})) \prec \sigma(u) \prec {\cal
I}^{\alpha}(g(a_{j+1}))$, then $g(a_{j}) \leq y \leq g(a_{j+1})$
by Proposition \ref{prec}, and there exists a unique $x \in V_{j}$
with $g(x)=y$. Thus $x \in V_{u} \neq 0$. Similarly we proceed if
$j \geq m$. If $ºuº$ is infinite, $v \sqsubseteq u$ and $v \in
2^{*}$, then $v$ satisfies the condition  of the Proposition, so
$V_{v} \neq 0$. Thus $V_{u} = \cap \{V_{v}º v \sqsubseteq u, v \in
A^{*}\}$ is nonempty by compactness. $\Box$ \mez

\begin{pro} \label{Law}
Let $w \in A^{N}$ be eventually periodic, and $j \in A$. If $w \in
L^{j}(w)$, then $L^{j}(w) \cap A^{*}$ is regular. If $w \in
L_{j}(w)$, then $L_{j}(w) \cap A^{*}$ is regular.
\end{pro}
Proof: For $u,v \in A^{*}$ define $u \approx v$ iff $(\forall s \in
A^{*})(us \in L^{j}(w) \Leftrightarrow vs \in L^{j}(w))$. By Nerode
Theorem (see \cite{kn:Hopcroft}) it suffices to prove, that
$\approx$ has a finite number of equivalence classes. If $u_{i}=j$,
and $u_{k}\neq j$ for $k<i$, then $u \approx \sigma^{i}(u)$. Thus
it suffices to investigate words which begin with $j$. If
$w=\overline{w_{ºn}}$ is periodic even, then for all $u \in A^{*}$,
$jw_{º2n}u \approx jw_{ºn}u$, since for all $i$ not divisible by
$n$, $w_{i-1}=j$ implies $\sigma^{i}(w)_{ºn} \prec w$, and
$w_{º2n}u \preceq w$ iff
$w_{ºn}u \preceq w$. Thus there is at most $2n$ equivalence
classes. If $w=\overline{w_{ºn}}$ is periodic odd, then $jw_{º3n}u
\approx jw_{ºn}u$, so there is at most $3n$ equivalence classes.
If $w$ is preperiodic, then we can write
$w=w_{0}...w_{m-1}\overline{w_{m}...w_{m+n-1}}$, where
$w_{m}...w_{m+n-1}$ is even. Since the set $\{ \sigma^{i}(w)º i \in
N\}$ is finite, there exists an integer $p$ such that
$\sigma^{i+1}(w)_{ºp} \prec w_{ºp}$ whenever $w_{i}=j$. Let $k$ be
an integer with $nk>p$ and put $r=m+nk$, $s=m+(n+1)k$. Then
$jw_{ºr}u \approx jw_{ºs}u$, so there is at most $s$ equivalence
classes. $\Box$

\begin{thm} \label{uniregular}
${\cal L}(I,g,\alpha)$ is a regular (recursive) language iff  all
${\cal I}_{\alpha}(g(a_{j}))$ and ${\cal I}^{\alpha}(g(a_{j}))$
are eventually periodic (recursive) sequences.
\end{thm}
Proof: If the condition is satisfied, then $L={\cal L}(I,g,\alpha)$
is regular (recursive) since it is an intersection of finite number
of regular (recursive) languages
(Propositions \ref{unilanguage} and \ref{Law}). To prove the
converse, suppose that $w={\cal I}^{\alpha}(g(a_{j}))$ is aperiodic
and $L$ is regular. Assume $j \geq m$. Since $a_{j} \in V_{jw}$,
 $jw_{ºk} \in L$ for any $k>0$. By Nerode Theorem   there exists
a right congruence $\sim$ of finite index on $A^{*}$,  such that
$L$ is a union of its
equivalence classes. There exists an infinite set $M \subseteq N$
such all $\{w_{ºm} º m \in M \}$ have the same parity. Assume they
are all even. Since $\sim$ is of finite index, there exist $m,n \in
M$, $m \neq n$, such that $jw_{ºm} \sim jw_{ºn}$. Since $\sim$ is
a right congruence, $jw_{º(m+k)} \sim jw_{ºn} \sigma^{m}(w)_{ºk}
\in L$ for any $k$, so $w_{ºn}\sigma^{m}(w)_{ºk} \preceq w$, and
$\sigma^{m}(w)_{ºk} \preceq \sigma^{n}(w)_{ºk}$, since $w_{ºn}$ is
even.  Interchanging $m$ and $n$ in the above argument we get
$\sigma^{n}(w)_{ºk} \preceq \sigma^{m}(w)_{ºk}$ for any $k$, and
therefore $\sigma^{n}(w) = \sigma^{m}(w)$.  This is, however, in
contradiction with the aperiodicity of $w$. If all $\{w_{ºm}º m \in
M\}$ are odd, or if $j<m$, the proof is similar. For the case of
recursivity suppose that $w={\cal I}^{\alpha}(g(a_{j}))$ is
nonrecursive and $L$ is  recursive. Again  $jw_{ºk} \in L$ for any
$k$. If $j \geq m$, then $w_{k} = \max \{iº jw_{ºk}i \in L\}$. If
$j < m$, then $w_{k} = \min \{iº jw_{ºk}i \in L\}$. Since $L$ is
recursive,  we get a recursive procedure for the computation of
$w$, which is a contradiction. $\Box$

\begin{dfn}
Denote $\alpha_{g} = \{›a,c!,›c,b!\}$ the standard cover of a
unimodal system $(I,g)$ with endpoints $a$, $b$, and turning point
$c$. The itinerary ${\cal I}(x) \in \overline{2^{*}}$ of $x \in I$
is defined by
${\cal I}(x)_{i} = 0\; \Leftrightarrow \;g^{i}(x)<c,\;\;\;$ ${\cal
I}(x)_{i} = 1\; \Leftrightarrow \;g^{i}(x)>c$ \\
for $0 \leq i< º{\cal I}(x)º = \inf\{n \geq 0 º g^{n}(x) = c \}$.
The kneading sequence of $(I,g)$ is ${\cal K}(g) = {\cal
I}(g(c))$. Denote also ${\cal I}_{g}(x) = {\cal
I}_{\alpha_{g}}(x)$, ${\cal I}^{g}(x) = {\cal I}^{\alpha_{g}}(x)$,
${\cal K}_{g} = {\cal I}_{g}(g(c))$, ${\cal K}^{g} = {\cal
I}^{g}(g(c))$.
\end{dfn}

\noindent
Thus ${\cal I}(x)$ is finite iff the orbit of $x$ contains
$c$, e.g. ${\cal I}(c) = \lambda$ is the empty word. If $g(c)=c$,
then ${\cal K}(g)=\lambda$. In general
\› \begin{array}{llll}
{\cal K}_{g} = {\cal K}(g), &
{\cal K}^{g} = {\cal K}(g)  &
         \mbox{if} & {\cal K}(g) \mbox{ is infinite},\\
{\cal K}_{g} = \overline{{\cal K}(g)1}, &
{\cal K}^{g} = {\cal K}(g)0\overline{{\cal K}(g)1}  &
         \mbox{if} & {\cal K}(g) \mbox{ is finite odd},\\
{\cal K}_{g} = \overline{{\cal K}(g)0}, &
{\cal K}^{g} = {\cal K}(g)1\overline{{\cal K}(g)0}  &
         \mbox{if} & {\cal K}(g) \mbox{ is finite even}\\
{\cal I}_{g}(x) = {\cal I}(x), &
{\cal I}^{g}(x) = {\cal I}(x)  &
         \mbox{if} & {\cal I}(x) \mbox{ is infinite},\\
{\cal I}_{g}(x) = {\cal I}(x)1{\cal K}_{g}, &
{\cal I}^{g}(x) = {\cal I}(x)0{\cal K}_{g} &
         \mbox{if} & {\cal I}(x) \mbox{ is finite odd},\\
{\cal I}_{g}(x) = {\cal I}(x)0{\cal K}_{g}, &
{\cal I}^{g}(x) = {\cal I}(x)1{\cal K}_{g} &
         \mbox{if} & {\cal I}(x) \mbox{ is finite even}.
\end{array} \!

\begin{dfn}
For $w \in \overline{2^{*}}$ denote $\Sigma_{w} = \{ u \in 2^{N}º
(\forall i >0)(\sigma^{i}(u) \preceq w)\} = L^{0}(w) \cap L^{1}(w)
\cap 2^{N}$. We say that $w$ is maximal, if $(\forall i)(\sigma^{i}(w)
\preceq w)$.
\end{dfn}

\begin{thm} \label{regularstandard}
${\cal L}(I,g,\alpha_{g})$ = ${\cal L}(\Sigma_{{\cal K}^{g}})$ is regular
iff ${\cal K}(g)$ is either finite or eventually periodic.
\end{thm}
Proof: Proposition \ref{unilanguage} and Theorem \ref{uniregular}.

\begin{pro} \label{lowknead}
$\Sigma_{{\cal K}{g}} = \{{\cal I}_{g}(x), {\cal I}^{g}(x)º x \in I \}$.
\end{pro}
Proof: If $x \in I$ and $i>0$, then either
$\sigma^{i}({\cal I}_{g}(x)) = {\cal I}_{g}(g^{i}(x))$, or
$\sigma^{i}({\cal I}_{g}(x)) = {\cal I}^{g}(g^{i}(x))$.
Since $g^{i}(x) \leq g(c)$, $\sigma^{i}({\cal I}_{g}(x))
\preceq {\cal K}_{g}$ by Proposition \ref{uplow}. Similarly for
$\sigma^{i}({\cal I}^{g}(x))$. If $u \in \Sigma_{{\cal K}_{g}}$,
then $V_{u} \neq 0$ by Proposition \ref{Law}. If there exists
$x \in V_{u}$, whose orbit does not contain $c$, then $u={\cal I}(x)$.
If $g^{i}(x)=c$, then $\sigma^{i+1}(u) = {\cal K}_{g}$, so either
$u = {\cal I}_{g}(x)$, or $u = {\cal I}^{g}(x)$. $\Box$ \mez

The first nonperiodic kneading sequence can be obtained by
successive applications of the doubling operator. The double of $s
\in 2^{*}-\{\lambda\}$ is ${\cal D}(s) = s\hat{s}$, where $\hat{s}$
is the sequence obtained from $s$ by changing the last bit. The
doubling operator can be iterated, and since $s \sqsubseteq {\cal
D}(s)$, ${\cal D}^{\infty}(s) =
\lim_{i \rightarrow \infty} {\cal D}^{i}(s)$ is well defined.  In
particular ${\cal D}^{\infty}(1)$ = $1011\ 1010 ...$ .

\begin{thm}
A unimodal system $(I,g)$ is regular if ${\cal K}_{g} \prec {\cal
D}^{\infty}(1)$, and nonrecursive otherwise.
\end{thm}
Proof: If ${\cal K}_{g} \prec {\cal D}^{\infty}(1)$, then the
itinerary of any point $x \in I$ is either finite or eventually
periodic. Thus if $\alpha$ is an
interval cover with the turning point among its cut points, then
all ${\cal I}^{\alpha}(g(a_{j}))$ and ${\cal I}_{\alpha}(g(a_{j}))$
are eventually periodic, so ${\cal L}(I,g,\alpha)$ is regular by
Theorem \ref{uniregular}. If $\alpha$ is an arbitrary interval
cover, then there exists a finer interval cover with cut point $c$,
and we get the result by Proposition \ref{refinement}.
On the other hand suppose that ${\cal K}(g) \succeq {\cal
D}^{\infty}(1)$. For a sequence $n_{i}$ of positive integers denote
$w = 1^{n_{0}}(10)^{n_{1}}(1011)^{n_{2}}...({\cal
D}^{i}(1))^{n_{i}}...$. Then $w \in {\cal L}(I,g,\alpha_{g})$, and
there exists point $x$ with ${\cal I}(x)=w$. If the sequence
$n_{j}$ is not recursive, then neither is $w$. $\Box$ \mez

Thus regularity and recursivity differentiate unimodal systems
similarly as topological entropy, since the topological entropy of
$(I,g)$ is zero iff ${\cal K}_{g} \preceq {\cal D}^{\infty}(1)$
(see \cite{kn:Milnor}).

\section{$S$-unimodal systems}

\begin{dfn}
A unimodal system $(I,g)$ is $S$-unimodal, if it has continuous
third derivation, $g''(c)<0$, and $Sg(x)<0$ for all $x \neq c$,
where $Sg(x) = \frac{g'''(x)}{g'(x)} -
\frac{3}{2}(\frac{g''(x)}{g'(x)})^{2}$ is the Schwarzian
derivative. \end{dfn}

\begin{thm} \label{guckenheimer}
Let $(I,g)$ be $S$-unimodal system. If ${\cal K}(g)$ is neither
finite nor periodic, then $V_{u}$ has empty interior for any $u \in
2^{N}$. If ${\cal K}(g)$ is either finite or periodic, and $u \in
2^{N}$, then $V_{u}$ has nonempty interior iff $u \in \Sigma_{{\cal
K}_{g}}$ and there exists $k \geq 0$ with ${\cal K}(g) \sqsubseteq
\sigma^{k}(u)$. \end{thm}
See Guckenheimer \cite{kn:Guckenheimer} or Collet and Eckmann
\cite{kn:Collet} for a proof. \mez

\begin{pro} \label{psi}
If $(I,g)$ is $S$-unimodal, then there exists a homomorphism $\psi:
(\Sigma_{{\cal K}_{g}},\sigma) \rightarrow (I,g)$, with $(\forall
u \in \Sigma_{{\cal K}_{g}})(\psi(u) \in V_{u})$, which maps
eventually periodic  points to eventually periodic points with the
same period.
\end{pro}
Proof: If ${\cal K}(g)$ is neither finite nor periodic, then by
Theorem \ref{guckenheimer}, $V_{u}$ is a singleton for each   $u
\in \Sigma_{{\cal K}_{g}}$, so the condition $\psi(u) \in V_{u}$
defines $\psi$ uniquely, and $\psi$ is clearly continuous.
By Proposition \ref{lowknead}, it is surjective.
Suppose now that ${\cal K}(g)$ = $\overline{w}$ is
periodic odd, and denote $n=ºwº$. Then $g^{n}$ is decreasing on
$V_{\overline{w}}$ and has therefore a unique $g^{n}$-fixed point
$\psi(\overline{w}) \in V_{\overline{w}}$. If $\sigma^{i}(u) =
\overline{w}$, then there is a unique $\psi(u) \in V_{u}$ with
$g^{i} \psi(u) = \psi(\overline{w})$. Thus $\psi$ is first defined
on the orbit of $\overline{w}$, and then on its preimages. So
defined $\psi$ is continuous,  since each $u$ for which
$\sigma^{i}(u) =\overline{w}$, is an isolated point. Indeed if $wwu
\in \Sigma_{{\cal K}_{g}}$, then $\sigma^{2n}(wwu) = u \preceq
\overline{w}$ by definition, and $\sigma^{n}(wwu) = wu \preceq
\overline{w}$, so  $u \succeq \sigma^{n}(\overline{w}) =
\overline{w}$, since $w$ is odd. Thus $u=\overline{w}$, and
$\overline{w}$ is an isolated point of $\Sigma_{{\cal K}_{g}}$.
If $\sigma^{i}(u) = \overline{w}$, then $u$ is isolated too,  since
$\sigma$ is continuous. Finally we assume that ${\cal K}(g)$
is either finite or periodic even. Then ${\cal K}_{g} =
\overline{w}$ is periodic even too. If $w=0$, then  $\Sigma_{{\cal
K}_{g}} = \{\overline{0},1\overline{0}\}$, and the
Proposition is trivial. If $w \neq 0$, then $10 \sqsubseteq w$, and
$w^{p}\overline{1} \in \Sigma_{{\cal K}_{g}}$ for any $p$. Thus
$\overline{w}$ is a limit point in this case and $V_{\overline{w}}
\neq V_{u}$ whenever $u$ is an initial substring of $\overline{w}$.
Denote $W = \cap \{ \overline{V_{u} - V_{\overline{w}}} º u \in
2^{*}, u \sqsubseteq \overline{w}\}$. Then $W$ is a nonempty
invariant set, and does not contain any inner point of
$V_{\overline{w}}$. Suppose that $W$ contains both endpoints of
$V_{\overline{w}}$. Then they would be unstable fixed points for
$g^{2n}$, and therefore there would exist a stable fixed point
between them. This is, however, in contradiction with $Sg <0$. Thus
$W$ consists of exactly one periodic point with period $n$ and we
define $\psi(\overline{w}) \in W$. The construction then proceeds
as  in the preceding case. It follows from the construction that
$\psi$ is continuous at $\overline{w}$, and also at all its
preimages. $\Box$

\begin{thm} \label{nonper}
Let $(I,g)$ be S-unimodal system such that ${\cal K}(g)$ is neither
finite nor periodic. Then $(I,g)$ is a factor of
$(\Sigma_{{\cal K}_{g}},\sigma)$.
\end{thm}
Proof: Theorem \ref{guckenheimer} and Proposition \ref{psi}

\begin{thm}
If $(I,g)$ is $S$-unimodal and ${\cal K}(g)$ is preperiodic, then
$(I,g)$ is a factor of a finite automaton.
\end{thm}
Proof: Theorems \ref{regularstandard}, \ref{regularshift} and
\ref{nonper}.

\begin{pro} \label{monotone}
Let $g: I \rightarrow I$ be a monotone (i.e. either nondecreasing
or nonincreasing) continuous map on a compact real interval $I$.
Then $(I,g)$ is a factor of a finite automaton.
\end{pro}
See \cite{kn:Kurka} for a proof.

\begin{thm}
If $(I,g)$ is $S$-unimodal and ${\cal K}(g)$ is either finite or
periodic, then $(I,g)$ is a factor of a finite automaton.
\end{thm}
Proof: Denote ${\cal K}_{g} = \overline{w}$, $ºwº=n$.
Consider an alphabet $A=\{0,1,2,3,4,5\}$, a projection $\nu : A
\rightarrow \{0,1\}$ defined by $\nu(a) = a$ mod 2, and a subshift
$\Sigma \subseteq A^{N}$ defined by $x \in \Sigma$ iff either $x
\in \Sigma_{\overline{w}}$, or $x=uv$, where $u,v$ satisfy
following conditions: $u \in \{0,1\}^{*}$, $v \in \{2,...,5\}^{N}$,
$\nu(uv) \in \Sigma_{\overline{w}}$, $w_{n-1}w_{0}...w_{n-2}$ is
not a final segment of $u$, $\nu\sigma(v) = \overline{w}$, and
$v_{i}\in \{4,5\}$ iff $n$ divides $i$. Since
$\Sigma_{\overline{w}}$ is regular, $\Sigma$ is regular too, and
there is a factorization $\varphi' : (X',f') \rightarrow
(\Sigma,\sigma)$. By Proposition \ref{psi} there is a
homomorphism $\psi : (\Sigma_{\overline{w}},\sigma) \rightarrow
(I,g)$. Thus we have a homomorphism $\psi \nu \varphi': (X',f')
\rightarrow  (I,g)$, which is, however, not surjective.  By
Proposition \ref{monotone} there is a factorization
$\varphi'':(X'',f'') \rightarrow (V_{\overline{w}},g^{n})$.  We
construct a finite automaton $(X,f)$, where $X = X' \times X''$,
$f(u',u'') = (f'(u'),f''(u''))$ if $\varphi'(u')_{0} \in \{4,5\}$
(this depends only on $u'_{0}$ and $u''_{0}$), and
$f(u',u'') = (f'(u'),u'')$ otherwise.
Then  $\sigma \nu \varphi'(u') = \overline{w}$ provided
$\varphi'(u')_{0} \in \{4,5\}$ . Define $\varphi: X \rightarrow I$
by
\› \begin{array}{lll}
\varphi(u',u'') = \psi \varphi'(u') & \mbox{ if } &
    \varphi'(u') \in 2^{N} \\
\varphi(u',u'') = g_{\nu \varphi'(u')_{0}}g^{n}\varphi''(u'')  &
\mbox{ if } &     \varphi'(u')_{0} \in \{4,5\} \\
\varphi(u',u'') = g_{\nu \varphi'(u')_{0}} \varphi(f'(u'),u'')  &
\mbox{ if } & \varphi(f'(u'),u'') \mbox{ has been defined}
\end{array} \!
(Here $g_{i}: ›a,g(c)! \rightarrow V_{i}$, are the two inverses of
$g$.) Then $\varphi(u',u'') \in V_{\nu\varphi'(u')}$, and
$\varphi$ is a surjection.  If $\varphi'(u')_{0} \in \{4,5\}$,
then  $\varphi f(u',u'')$ = $\varphi(f'(u'),f''(u''))$ =
$g_{\nu\varphi'f'(u')_{0}} \varphi((f')^{2}(u'),f''(u''))$ = ...
$g_{\nu\varphi'(u')_{1}} ... g_{\nu\varphi'(u')_{n-1}}
\varphi((f')^{n}(u'),f''(u''))$ =
$g_{\nu\varphi'(u')_{1}} ... g_{\nu\varphi'(u')_{n}} g^{n}
\varphi'' f''(u'')$ = $\varphi'' f''(u'')$ = $g^{n}
\varphi''(u'')$ =  $g \varphi(u',u'')$. If $\varphi'(u')$ is
eventually periodic, and if $\varphi'(u')_{0} \not\in \{4,5\}$,
then $\varphi f(u',u'') = \varphi(f'(u'),u'') = g \varphi(u',u'')$.
Thus $\varphi$ is a factorization. $\Box$

\begin{cor}
An $S$-unimodal system with either finite or eventually periodic
kneading sequence is a factor of a finite automaton. Thus
$S$-unimodal systems with preperiodic kneading sequences are
factors of finite automata, which are not regular.
\end{cor}

\begin{exm}
If ${\cal K}(g) = {\cal D}^{\infty}(1)$, then $(I,g)$ is not a
factor of a finite automaton.
\end{exm}
Proof: For any $k>0$ there exists an open invariant neighbourhood
of $\omega(c)$, which does not contain periodic points with periods
less than $2^{k}$. By Theorem \ref{sizeper}, $(I,g)$ is not a
factor of a finite automaton (see \cite{kn:Kurka} for details).
\mez

We conjecture that an $S$-unimodal system with aperiodic kneading
sequence is not a factor of a finite automaton.

\end{document}